\documentclass[fleqn,usenatbib]{mnras}

\usepackage[T1]{fontenc}
\usepackage{ae,aecompl}

\usepackage{graphicx}	
\usepackage{amsmath}	
\usepackage{amssymb}	

\newcommand{\simgt}{\,\rlap{\lower 3.5 pt \hbox{$\mathchar \sim$}} \raise
1pt \hbox {$>$}\,}
\newcommand{\simlt}{\,\rlap{\lower 3.5 pt \hbox{$\mathchar \sim$}} \raise
1pt \hbox {$<$}\,}

\newcommand{\nh}{\langle n_\textrm{H}\rangle}
\newcommand{\nion}{\dot{N}_\textrm{ion}}
\newcommand{\nionbf}{\dot{\mathbf{N}}_\textbf{ion}}
\newcommand{\trec}{t_\textrm{rec}}
\newcommand{\fesc}{f_\textrm{esc}}

\newcommand{\Ob}{\Omega_\textrm{b}}
\newcommand{\Om}{\Omega_\textrm{m}}
\newcommand{\OL}{\Omega_\Lambda}

\newcommand{\lya}{Ly$\alpha$}

\newcommand{\xHI}{{\overline{x}_\textrm{{\textsc{hi}}}}}
\newcommand{\xHII}{{\overline{x}_\textrm{{\textsc{hii}}}}}

\newcommand{\MUV}{M_\textsc{uv}}

\newcommand{\kms}{km s$^{-1}$}

\newcommand{\JWST}{\textit{JWST}}

\newcommand{\BE}{\begin{equation}}
\newcommand{\EE}{\end{equation}}
\newcommand{\BEA}{\begin{eqnarray}}
\newcommand{\EEA}{\end{eqnarray}}

\defcitealias{PlanckCollaboration2018}{P18} 	  	

\title[Non-parametric $\nion(z)$]{Model-independent constraints on the hydrogen-ionizing emissivity at $z>6$}

\author[C. A. Mason et al.]{Charlotte A. Mason$^{1}$\thanks{E-mail: charlotte.mason@cfa.harvard.edu}\thanks{Hubble Fellow},
Rohan P. Naidu$^{1}$, Sandro Tacchella$^{1}$ and Joel Leja$^{1}$
\\
$^{1}$Center for Astrophysics \,|\, Harvard \& Smithsonian, 60 Garden St, Cambridge, MA, 02138, USA
}

\date{Accepted XXX. Received YYY; in original form ZZZ}

\pubyear{2019}

\begin{document}
\label{firstpage}
\pagerange{\pageref{firstpage}--\pageref{lastpage}}
\maketitle

\begin{abstract}
%
Modelling reionization often requires significant assumptions about the properties of ionizing sources. Here, we infer the total output of hydrogen-ionizing photons (the ionizing emissivity, $\nion$) at $z=4-14$ from current reionization constraints, being maximally agnostic to the properties of ionizing sources. We use a Bayesian analysis to fit for a non-parametric form of $\nion$, allowing us to flexibly explore the entire prior volume.
We infer a declining $\nion$ with redshift at $z>6$, which can be used as a benchmark for reionization models. Model-independent reionization constraints from the CMB optical depth and \lya\ and Ly$\beta$ forest dark pixel fraction produce $\nion$ evolution ($d\log_{10}\nionbf/dz|_{z=6\rightarrow8} = -0.31\pm0.35$\,dex) consistent with the declining UV luminosity density of galaxies, assuming constant ionizing photon escape fraction and efficiency. 
Including measurements from \lya\ damping of galaxies and quasars produces a more rapid decline: $d\log_{10}\nionbf/dz|_{z=6\rightarrow8} =-0.44\pm0.22$\,dex, steeper than the declining galaxy luminosity density (if extrapolated beyond $\MUV \simgt -13$), and constrains the mid-point of reionization to $z = 6.93\pm0.14$.
\end{abstract}




\section{Introduction}
\label{sec:intro}
The ultra-violet (UV) radiation of the first populations of stars and accreting black holes reionized hydrogen in the intergalactic medium (IGM) in the Universe's first billion years. Understanding the reionization process enables us to learn about the properties of these first light sources \citep[e.g.,][]{Loeb2001,Robertson2010}. 

However, to model reionization and interpret observations, significant assumptions are often made about the properties of these sources. Were they galaxies or quasars? What was their number density? How clustered were they? How hard were their ionizing spectra? A common approach, assuming reionization is driven by galaxies, is to parameterise the comoving density of ionizing photons, $\nion$, as a product of: the galaxy UV luminosity density, $\rho_L$; a conversion from observed UV photons (rest-frame $\sim 1500$\,\AA) to unobserved hydrogen-ionizing photons ($< 912$\,\AA), $\xi_\mathrm{ion}$; and the fraction of photons which escape galaxies to ionize the IGM, $\fesc$ \citep[e.g.,][]{Robertson2010}. However, in the reionization epoch, $z\simgt6$, $\fesc$ cannot be directly measured due to the high opacity of the IGM to ionizing photons \citep[e.g.,][]{Madau1995}, and the most straightforward way to estimate $\xi_\mathrm{ion}$ requires dust-corrected H$\alpha$ emission \citep{Leitherer1995} which calls for $\simgt4.6\,\mu$m spectroscopy e.g. with \JWST\ NIRSpec. Thus these parameters are often treated as constants in models \citep[e.g.,][]{Robertson2015,Bouwens2015}. One may ask how valid such assumptions are in the context of current constraints on reionization.

In the past year, several independent measurements provided evidence that reionization was a late ($z<9$), moderately extended ($\Delta z \simlt 4$), process. The optical depth to cosmic microwave background (CMB) photons provides an integral constraint: the \citet[][hereafter P18]{PlanckCollaboration2018} results suggest reionization's midpoint is at $z=7.7\pm0.7$. Observations of reduced Lyman-alpha (\lya) emission from high-redshift sources can be used to measure the neutral content of the IGM \citep[e.g.,][]{McQuinn2007a,Dijkstra2011,Mason2018}, as \lya\ photons are absorbed by neutral hydrogen. Damped \lya\ emission from galaxies and quasars currently provide the most precise constraints on the timeline of reionization, with most observations indicating a substantially neutral $z>7$ IGM \citep{Banados2017a,Davies2018,Mason2018,Mason2019c,Hoag2019a,Greig2019}.

The low number density of quasars observed at $z>6$ \citep[e.g.,][]{Parsa2018} has motivated recent work on galaxy-driven reionization. Fixing $\fesc=0.2$ and $\xi_\mathrm{ion} = 10^{25.2}$\,Hz/erg, \citet{Bouwens2015} and \citet{Robertson2015} found $z<10$ galaxies could reionize the IGM by $z\sim6$, providing undetected faint galaxies contribute significantly (extrapolating the galaxy luminosity function (LF) to $\MUV < -13$ with steep LF faint-end slopes, $\alpha \simlt -2$). However, these studies used parametric models, which could rule out possible evolutionary pathways \textit{a priori}: \citet{Bouwens2015} parameterised $\log_{10}{\nion} \propto z$; and \citet{Robertson2015} modelled the cosmic star formation rate (SFR) density in a parametric form based on $z\simlt8$ measurements.

Other works have explored models for $\fesc$ and $\xi_\mathrm{ion}$, allowing them to vary with galaxy properties. \citet{Finkelstein2019} used $\fesc$ as a function of halo mass, obtained in hydrodynamical simulations, where the lowest mass galaxies have the highest escape fractions, and modelled the galaxy UV luminosity density based on extrapolating $z\leq10$ UV LF fits. By shifting the load of reionization to the lowest mass galaxies \citet{Finkelstein2019} find a more extended reionization process and relatively flat $\nion(z)$, and also require a contribution from quasars at the end of reionization. \citet{Duncan2015} modelled the evolution of $\fesc$ and $\xi_\mathrm{ion}$ as functions of UV luminosity and spectral slope $\beta$ based on stellar population modelling, aiming to reproduce the $\nion(z)$ inferred by \citet{Bouwens2015}. \citet{Duncan2015} predict that faint blue galaxies produce and emit the most ionizing photons, which softens the requirement to extrapolate the LF in their work. \citet{Naidu2019}, based on the empirical galaxy evolution model of \citet{Tacchella2018a}, model $\fesc$ as a function of SFR surface density \citep[see also,][]{Sharma2017b,Seiler2019}, motivated by observations of Lyman Continuum leakers, and $\xi_\textrm{ion}$ using stellar population synthesis. With this model, \citet{Naidu2019} find $M_{\mathrm{UV}}<-18$ galaxies contribute $>80\%$ of the reionization budget \citep[see also,][]{Sharma2018}. However, \citet{Greig2017c} demonstrated that significant degeneracies exist between reionization parameters, which could not be broken with current reionization constraints, motivating an integrated approach to modelling the timeline and sources of reionization.

In this work, we seek to update our knowledge of $\nion(z\simgt5)$, by constraining it from recent estimates of the reionization timeline, and ask how consistent simple parametric models are with the allowed form of $\nion(z)$. This method enables us to estimate $\nion(z)$ at redshifts higher than is possible with the \lya\ forest, which becomes too heavily absorbed at $z>6$ \citep[e.g.,][]{Fan2006,Becker2013}. We aim to be maximally agnostic to the evolution of $\nion$ and so in a novel step we fit for a \textit{non-parametric} form of $\nion(z)$ allowing us to flexibly explore the allowed space. This offers an advantage over parametric models which rule out physically possible evolution \textit{a priori}.

The paper is structured as follows: Section~\ref{sec:reion} describes the reionization history model; Section~\ref{sec:nion} describes the method for inferring the redshift evolution of $\nion$; Section~\ref{sec:results} presents our inferred $\nion(z)$ and reionization history. We discuss our results in Section~\ref{sec:disc}, and summarise in Section~\ref{sec:conc}.

We use the \citet{PlanckCollaboration2015} cosmology: $(\OL, \Om, \Ob, n,  \sigma_8, H_0) =$ (0.69, 0.31, 0.048, 0.97, 0.81, 68 \kms\ Mpc$^{-1}$), as implemented in \verb|astropy|. Magnitudes are in the AB system. Distances, volumes, and densities are comoving. Confidence levels are 68\% unless otherwise stated.

\section{Modelling the reionization history}
\label{sec:reion}

Reionization progresses as ionizations overtake recombinations in the IGM. The reionization history: the volume-averaged ionized hydrogen fraction as a function of redshift, $\xHII(z)$, can be calculated by solving the following differential equation ~\citep[e.g.,][]{Madau1999}:
\BE  \label{eqn:reion_Q}
	\frac{\mathrm{d}\xHII}{\mathrm{d}t} = \frac{\nion}{\nh} - \frac{\xHII}{\trec}
\EE
where $\nion$ is the number density of ionizing photons in Mpc$^{-3}$ s$^{-1}$ and $\nh$ is the mean number density of hydrogen atoms. The recombination time of the IGM is 
$\trec(z) = \left[C\alpha_B(T)n_e(1+z)^3 \right]^{-1}$,
where $\alpha_B(T)$ is the case B hydrogen recombination (i.e. opaque IGM) coefficient, $n_e$ is the number density of electrons (assuming singly ionized He), and $C=\langle n_H^2 \rangle/\nh^2$ is the `clumping factor' which accounts for inhomogeneity in the IGM \citep{Madau1999}. For computational efficiency we fix $C=3$, motivated by IGM modelling and simulations \citep[e.g.,][]{Shull2012,Finlator2012,Kaurov2015}.  Appendix~\ref{app:clumping} discusses how our results are insensitive to this assumption. Throughout most of this work we use the \textit{neutral} fraction $\xHI = 1 - \xHII$.

Assuming galaxies produce the bulk of ionizing photons during reionization $\nion$ can be expressed as: 
\BE \label{eqn:reion_nion_G}
    \dot{N}_\textrm{ion,G} = \fesc \xi_\textrm{ion} \rho_L
\EE
where $\fesc$ is the fraction of ionizing photons which escape galaxies to the IGM; $\xi_\textrm{ion}$ is the production rate of ionizing photons per UV luminosity, in Hz/ergs, which depends on the stellar populations' initial mass function, metallicity, age and dust content; and $\rho_L$ is the dust-corrected UV luminosity density \citep[e.g.,][]{Robertson2010}. These parameters are likely functions of at least mass and redshift, but $\fesc$ and $\xi_\textrm{ion}$ are commonly treated as constant for simplicity.

The ionizing photon production rate from quasars can be derived from their spectral energy distribution (SED):
\BE \label{eqn:reion_nion_Q}
    \dot{N}_\textrm{ion,Q} = f_\textrm{esc,Q} \int d\nu \, \frac{\epsilon_{\nu}}{h\nu}
\EE
where the integral limits are from $1-4$\,Ry (the energy of hydrogen-ionizing photons) and $f_\textrm{esc,Q}$ is the ionizing photon escape fraction from quasars (usually assumed $\sim1$). The UV SED of quasars is assumed to follow a double-power law, $\epsilon_\nu \propto \nu^\alpha$, with a pivot at 912\,\AA. For $<912$\,\AA\ we use $\alpha=-1.70$ \citep{Madau2015,Kulkarni2018}. This value was derived from $M_\textrm{UV} \sim -27$ $z\sim2$ quasars \citep{Lusso2015}. $\alpha$ may be steeper for lower luminosity quasars \citep[e.g., -0.56 -- -1.4,][]{Stevans2014,Scott2004}, which could increase the normalisation of $\nion$ by up to $\sim0.5$ dex.

\section{Non-parametric inference of $\nion$}
\label{sec:nion}

To be maximally agnostic about the form of $\nion$ we fit for it as a non-parametric function of redshift, i.e. $\nion(z)$ can take \textit{any} value at any redshift. We use redshift bins of width $\Delta z =1$, assuming smooth evolution on that scale, and fit for $\nionbf(\mathbf{z})$ at $\mathbf{z}= \begin{bmatrix} 4 \; \dots \; 14 \end{bmatrix}$, i.e. 11 parameters. In the following we use $\nion$ to refer to the ionizing emissivity in general and the vector notation $\nionbf$ to refer specifically to our inferred values of $\nion$ in each redshift bin. Using Bayes' theorem, the posterior probability for $\nionbf(\mathbf{z})$ is:
\BE \label{eqn:nion_post}
p(\nionbf \,|\, \textrm{data}) = \mathcal{L}(\textrm{data} \,|\, \nionbf)\, p(\nionbf)
\EE
where $\mathcal{L}(\textrm{data} \,|\, \nionbf)$ is the likelihood of obtaining observed data given our model $\nionbf$ (described in Section~\ref{sec:nion_like}). We set the prior, $p(\nionbf)$, such that:
\begin{itemize}
    \item $\log_{10}\nionbf(z=5)$ is uniformly distributed between 49 and 53 ($\nion$ in units of Mpc$^{-3}$ s$^{-1}$), motivated by measurements of $\nion$ at $z\simlt5$ \citep{Becker2013}.
    \item The gradient between redshift steps, $d\log_{10}\nionbf/dz$, is uniformly distributed between -1 and 1 dex. This is motivated by the SFR/luminosity density of galaxies and quasars which fall by $<1$ dex per redshift at these redshifts \citep[e.g.,][]{Bouwens2015b,Finkelstein2015a,Oesch2018,Kulkarni2018}, even if bright galaxies dominate reionization \citep{Sharma2018}. The luminosity density of observable galaxies falls only $\sim0.2$ dex per redshift at these redshift, thus we do not expect enormous jumps in $\nion(z)$.
\end{itemize}
Figure~\ref{fig:nion} shows the posterior is not restricted by this prior.

In each likelihood call we sample $\nionbf$, linearly interpolate it over a redshift grid $\Delta z=0.2$ and solve the reionization history Equation~\ref{eqn:reion_Q}.
While e.g. $\fesc$ in individual galaxies could fluctuate on these timescales due to supernova feedback \citep{Trebitsch2017}, the average over the ensemble galaxy population should be smooth. The likelihood for the data, given $\nionbf$, is described in the next section. 

To obtain the posterior distribution for $\nion(z)$ we use dynamic nested sampling via \verb|dynesty|\footnote{\url{https://dynesty.readthedocs.io/}} \citep{Speagle2019}, with the sampling settings optimised for posterior estimation.

\subsection{Measurements used in $\nion$ likelihood}
\label{sec:nion_like}

We use two sets of reionization constraints to create two likelihood functions. The total likelihood is the product of the individual likelihoods $\mathcal{L}_i$ for the below data given our model $\nionbf$. Following \citet{Greig2017c}, we define a `Gold' sample of model-independent measurements, and an additional sample of more model-dependent measurements (described below). The `Gold' sample constraints are:

\begin{enumerate}
    \item \textbf{CMB electron scattering optical depth.} CMB photons scatter off a fraction $\tau_\mathrm{es}$ of free electrons created during reionization, suppressing CMB anisotropies by $\exp(-\tau_\mathrm{es})$, below angular scales corresponding to the size of the cosmological horizon at reionization.
    \BE  \label{eqn:reion_tau}
    	\tau_\mathrm{es}(z) = \int_0^z dz' \; \sigma_\textsc{t} n_e (1+z')^2 \, \xHII(z') \frac{c}{H(z')} 
    \EE
    where $c$ is the speed of light, $\sigma_\textsc{t}$ is the Thomson scattering cross section, $n_e$ is the comoving number of free electrons \citep[assuming doubly-ionized Helium at $z<4$, following][the exact timing of Helium reionization has a negligible impact on our results]{Kuhlen2012,Bouwens2015}, and $H(z)$ is the Hubble parameter. As $\tau_\mathrm{es}$ is measured at $z\sim1100$ for the CMB it provides only an integral constraint on reionization and cannot give precise information about its timing.
    
    In each call of our likelihood we solve for $\tau_\mathrm{es}(\nionbf, z=15)$ (assuming fully neutral IGM) and calculate the likelihood for obtaining the \citetalias{PlanckCollaboration2018} optical depth: $\tau_\mathrm{es} = 0.054\pm 0.007$, from our model, assuming Gaussian uncertainties.

    \item \textbf{\lya\ forest dark fraction.} The fraction of dark pixels in the \lya\ and Ly$\beta$ forest of $z\sim6$ quasars (hereafter, `dark fraction') provides a model-independent constraint on $\xHI$, as neutral hydrogen in the IGM produces fully saturated absorption in the \lya\ and Ly$\beta$ forest \citep{Mesinger2010}. This is degenerate with absorption by self-shielded neutral gas post-reionization, so measurements provide upper limits. We use dark fraction measurements by \citet{McGreer2015}: $\xHI(z=5.6) < 0.04+0.05 \; (1\sigma)$, $\xHI(z=5.9) < 0.06+0.05$ and $\xHI(z=6.1) < 0.38+0.20$. Following \citet{Greig2017c} we implement these via a likelihood with a uniform probability for $\xHI < \xHI_\mathrm{lim}$ and a half-Gaussian distribution with $\mu=\xHI_\mathrm{,lim}$ and $\sigma=\sigma_\mathrm{lim}$ for higher values of $\xHI$.
\end{enumerate}

The second likelihood adds the following measurements of $\xHI(z)$ from \lya\ damping of high-redshift sources, which account for patchy reionization in statistically robust methods. $\xHI$ can be measured from an observed reduction in \lya\ as neutral hydrogen in the IGM absorbs \lya\ photons. However, such measurements rely on modelling the  sources observed as backlights to the neutral IGM:
\begin{enumerate}
    \item \textbf{\lya\ equivalent width (EW) distribution.} We use $\xHI$ constraints at $z\sim7,7.5$ and 8 from \citet{Mason2018,Hoag2019a,Mason2019c}. These measurements infer $\xHI$ from the redshift evolution of the \lya\ EW distribution in Lyman-break galaxies via comparison with realistic IGM simulations. As these constraints were relative to $z=6$ (i.e. assuming the IGM was fully ionized at $z=6$) we use the relative measurement $\Delta_{6-z}\xHI = \xHI(z) - \xHI(6)$ in our likelihood. We use the posteriors $p(\xHI(z_\mathrm{meas}))$ from those works to directly calculate the likelihood $p(\Delta_{6-z}\xHI \,|\, \nionbf)$ from our model. To account for the redshift uncertainty in these measurements due to using photometric samples ($\Delta z \sim 1$), we draw 1000 redshifts from the photometric redshift distribution for sources used in each measurement, maintaining detailed balance by using same drawn redshifts in every likelihood calculation, and use the median of the likelihoods calculated for the drawn redshifts.
    \item \textbf{\lya\ emitter clustering.} Reionization increases the clustering of \lya\ emitting galaxies. We use the constraint of $\xHI(z=6.6) \leq 0.5\,(1\sigma)$ by \citet{Sobacchi2015} obtained from comparing the angular correlation function of \lya\ emitters by \citet{Ouchi2010} to reionization simulations. We implement the likelihood as half-Gaussian distribution with $\mu=0$ and $\sigma=0.5$.
    \item \textbf{\lya\ damping wings in quasar spectra.} The neutral IGM creates smooth \lya\ damping absorption in quasar spectra. We use recent constraints on $\xHI$ from two $z>7$ quasars by \citet{Davies2018} and \citet{Greig2019}. We use the $\xHI$ posteriors from those works to calculate likelihoods for our model \citep[using the `Intermediate HII regions' IGM simulation from][]{Greig2019}. We note these works obtained slightly different constraints (though consistent within $1-1.5\sigma$) primarily due to differences in modelling the intrinsic quasar spectra. We include both constraints to `marginalise' over differences in modelling.
\end{enumerate}

These \lya\ damping $\xHI$ measurements are all obtained by comparison to large-scale reionizing IGM simulations. By construction these rely on parametric `sub-grid' models for the properties of reionizing sources, which set the timing and morphology of reionization. However, currently these $\xHI$ measurements are relatively insensitive to the properties of reionizing sources: there is little difference between $\xHI$ results obtained using different simulation setups \citep{Sobacchi2015,Mason2018,Greig2019}, meaning our results should be robust to the simulation assumptions. This is due to the structure of the reionizing IGM depending primarily on $\xHI$, with a lesser some dependence on the clustering scale of the primary reionizing sources, and only weak dependence on the redshift of reionization \citep[as the luminosity-weighted power spectra of galaxies does not change much $z\sim6-10$,][]{McQuinn2007}. The impact of reionization morphology is mostly diluted in the measurements from galaxies, which span a redshift range greater than typical size of ionizing bubbles ($<\Delta z = 0.1$), and in the case of quasars due to the large uncertainties in $\xHI$ measurements from single objects.

Additional constraints on reionization come from the patchy kinetic Sunyaev-Zel'dovich (kSZ) effect, which requires large 3D simulations to model, and is beyond the scope of this work. \citet{Greig2017c} demonstrate current patchy kSZ results \citep{George2015} favour late reionization, but do not provide more constraining power on $\xHI(z)$ than $\tau_\mathrm{es}$ and the dark fraction.

\section{Results}
\label{sec:results}

\begin{figure*}
    \includegraphics[width=\textwidth]{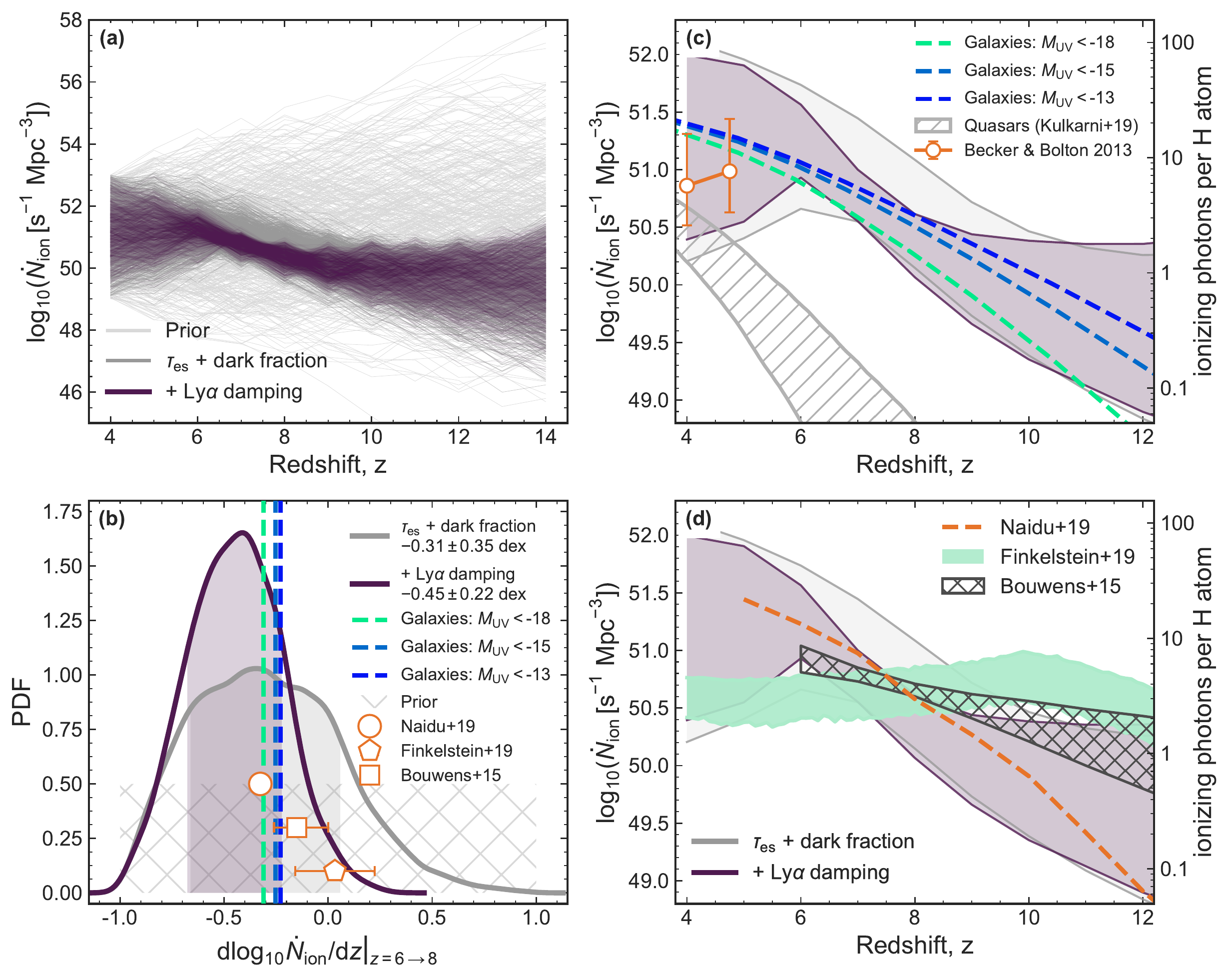}
    \vspace{-16pt}
    \caption{Redshift evolution of $\nion$. 
    \textbf{(a)} 
    1000 draws from our posteriors for $\nionbf$. Dark grey lines show $\nionbf$ inferred using only the \citetalias{PlanckCollaboration2018} optical depth, $\tau_\mathrm{es}$, and dark fraction constraints. dark purple lines show $\nionbf$ inferred using the additional $\xHI$ constraints from \lya\ damping. Light grey lines show 500 draws from the prior. The posteriors in both cases are substantially narrower than the prior, especially $5 < z < 10$, demonstrating that we have updated our knowledge on $\nionbf$.
    \textbf{(b)}
    The distribution of gradients in $\log_{10}\nion$ between $z=6$ and $z=8$. Colours same as above, shading show 68\% confidence region. Prior distribution shown as light grey cross-hatch. The grey diagonal hatch region shows range of the galaxies model described in panel (c). Orange points show models by \citet{Bouwens2015}, \citet{Finkelstein2019} and \citet{Naidu2019}.
    \textbf{(c)}
    Same as (a) but now shaded regions shows the 68\% confidence region for our inferred $\nionbf$. Circular points show the $\nion$ measurements by \citet{Becker2013}. We also plot the range of $\nion(z)$ allowed by the evolving number density of galaxies \citep[with constant $\fesc$ and $\xi_\mathrm{ion}$, integrating the UV LF down to $\MUV < -18, -15, -13$,][dashed lines]{Mason2015} and quasars \citep[constant $\fesc$ and EUV slope,][circular hatch]{Kulkarni2018} -- described in Section~\ref{sec:results_nion}.
    \textbf{(d)} 
    Same as (c), comparing $\nion$ models from \citet[][cross hatch]{Bouwens2015}, \citet[][light green region]{Finkelstein2019} and \citet[][orange dashed line]{Naidu2019}.}
    \label{fig:nion}
\end{figure*}

\subsection{Evolution of $\nion(z)$}
\label{sec:results_nion}

Figure~\ref{fig:nion} shows $\nionbf(\mathbf{z})$ inferred from the measurements described in Section~\ref{sec:nion_like}. Panel (a) shows draws from our posteriors and the prior, to demonstrate how $\nion$ evolution is constrained when the likelihood is included (Section~\ref{sec:nion_like}). In both cases the inferred $\nionbf$ is a smooth, declining function of redshift, with the tightest posteriors obtained at $z\sim6-8$ where the majority of the reionization history constraints are found (Figure~\ref{fig:history}). When the \lya\ damping $\xHI$ constraints are included, the posterior for $\nionbf$ is noticeably tighter and prefers a more rapid decline with redshift.

This is clearer in panel (b) which shows the distribution of derivatives of $\log_{10}\nionbf$ between $z=6\rightarrow8$ from our $\nionbf$ posteriors. $d\log_{10}\nionbf/dz|_{z=6\rightarrow8} = -0.31\pm0.35$ inferred from the $\tau_\mathrm{es}$ and dark fraction constraints, and $-0.44\pm0.22$ when the \lya\ damping constraints are added. We also compare to models by \citet{Bouwens2015}, \citet{Finkelstein2019} and \citet{Naidu2019}, also shown in panel (d) and described in more detail below.

Panel (c) compares our inferred $\nionbf$ to simple `Galaxies' and `Quasars' models. For the `Galaxies' model we take the galaxy UV luminosity density from the \citet{Mason2015} UV luminosity function (LF) model \citep[consistent with observations from $0 < z < 10$, see also,][]{Tacchella2013}, and calculate $\dot{N}_\textrm{ion,G}$ (Equation~\ref{eqn:reion_nion_G}), fixing $\fesc = 0.2$ and $\log_{10}(\xi_\mathrm{ion}) = 25.2$ for comparison with previous works \citep{Bouwens2015,Robertson2015}. To encompass uncertainty in the faint-end cut-off of the galaxy LF we show the range of $\dot{N}_\textrm{ion,G}$ allowed by integrating the LFs to $\MUV < -13$ \citep[comparable to the atomic cooling limit at $z\sim8-10$, $M_h\simgt10^8\,M_\odot$,][resulting in a flatter luminosity density and $\nion$]{Bromm2011}, $\MUV < -15$ \citep[the depth reached in deep HST imaging of galaxy cluster lensed fields, e.g.,][]{Livermore2017,Bouwens2017a} and $\MUV < -18$. For the quasar model we take the quasar ionizing emissivity at 912\,\AA\ from a homogenised compilation by \citet{Kulkarni2018} and assume a power-law quasar UV SED to calculate the hydrogen-ionizing emissivity (Equation~\ref{eqn:reion_nion_Q}). We assume the ionizing photon escape fraction from quasars, $f_\textrm{esc,Q}=1$, though this may be optimistic \citep[e.g.,][]{Micheva2017}, thus our model shows the upper limit of the quasar contribution. To encompass the uncertainty in the quasar LF faint-end cut-off we show the range of $\dot{N}_\textrm{ion,Q}$ obtained by integrating to $\MUV < -18$ or $\MUV < -21$. The galaxy model is consistent with $\nionbf$ inferred from $\tau_\mathrm{es}$ and the dark fraction constraints. However, including the \lya\ damping constraints requires a steeper evolution in $\nion$ between $z\sim6-8$, as described above and shown in panel (b). The majority of the posterior inferred for the slope $d\log_{10}\nionbf/dz|_{z=6\rightarrow8}$ favours steeper values than the galaxy luminosity density: 72\% of the posterior has $d\log_{10}\nionbf/dz|_{z=6\rightarrow8} < -0.31$\,dex (steeper than the slope of the galaxy luminosity density if $\MUV < -18$),  80\% of the posterior has $d\log_{10}\nionbf/dz|_{z=6\rightarrow8} < -0.25$\,dex, steeper than the galaxy luminosity density for $\MUV < -15$, and 83\% has $d\log_{10}\nionbf/dz|_{z=6\rightarrow8} < -0.23$\,dex (steeper than the galaxy luminosity density for $\MUV < -12$). The inferred steep $\nion(z)$ gradient thus makes a model with constant $\fesc$ and $\xi_\mathrm{ion}$ and/or abundant faint galaxies less likely.

Panel (d) compares our inferred $\nionbf$ to models by \citet{Bouwens2015}, \citet{Finkelstein2019} and \citet{Naidu2019}. The \citet{Bouwens2015} model was derived using older constraints on the reionization history, and used tighter priors on $\nion$, but is mostly consistent with our $\nionbf$ inferred from the \citetalias{PlanckCollaboration2018} $\tau_\mathrm{es}$ and dark fraction. This model is similar to the `Galaxies' model in panel (c), with its evolution consistent with the galaxy luminosity density, dropping $\sim0.2$ dex per redshift. It is difficult for such a model, where $\fesc$ and $\xi_\mathrm{ion}$ are constant, to match the $\nion(z)$ evolution inferred from the full set of $\xHI$ constraints which show later, more rapid, reionization -- $d\log_{10}\nionbf/dz|_{z=6\rightarrow8} = -0.44\pm0.22$ dex. 

The models by \citet{Finkelstein2019} and \citet{Naidu2019} model $\fesc$ as functions: of halo mass (with the highest $\fesc$ in low mass halos); and SFR surface density \citep[highest $\fesc$ in high SFR surface density objects, see also][]{Sharma2017b,Seiler2019} respectively. The \citet{Finkelstein2019} $\nion$ model slightly increases $z\sim4\rightarrow10$, due to low mass halos being more prevalent at higher redshifts with a steep LF faint-end slope ($\alpha < -2$). As demonstrated in panel (b), with the $\tau_\mathrm{es}$ and dark fraction constraints, 21\% of our $\nionbf$ posterior increases over $z=6\rightarrow8$ ($d\log_{10}\nionbf/dz|_{z=6\rightarrow8} > 0$), and only 3\% when the \lya\ damping constraints are included, making an increasing $\nion(z)$ much more unlikely. The \citet{Naidu2019} SFR surface-density $\fesc$ model shifts the burden of reionization onto brighter galaxies, producing a steeper rise in $\nion$ with decreasing redshift \citep[see also,][]{Sharma2018,Seiler2019}. We note that the \citet{Naidu2019} model was fit to most of the same constraints as presented here, so is likely to match by construction, but demonstrates a physically-motivated model which can match the inferred evolution in $\nion$.

The right axes of panels (c) and (d) show the number of ionizing photons per hydrogen atom released in a Hubble time: $\nion/\langle n_\mathrm{H} \rangle/H(z)$, calculated at $z=6$. We infer $\sim5-23$ (68\% range) ionizing photons per hydrogen atom from our $\nionbf$ posteriors including the \lya\ damping constraints ($\sim3-36$ with only $\tau_\mathrm{es}$ and the dark fraction), supporting a scenario where ionizing photons are relatively abundant in the late stages of reionization \citep{Becker2013}.

\subsection{Reionization history and CMB optical depth}
\label{sec:results_history}

\begin{figure}
	\includegraphics[width=\columnwidth]{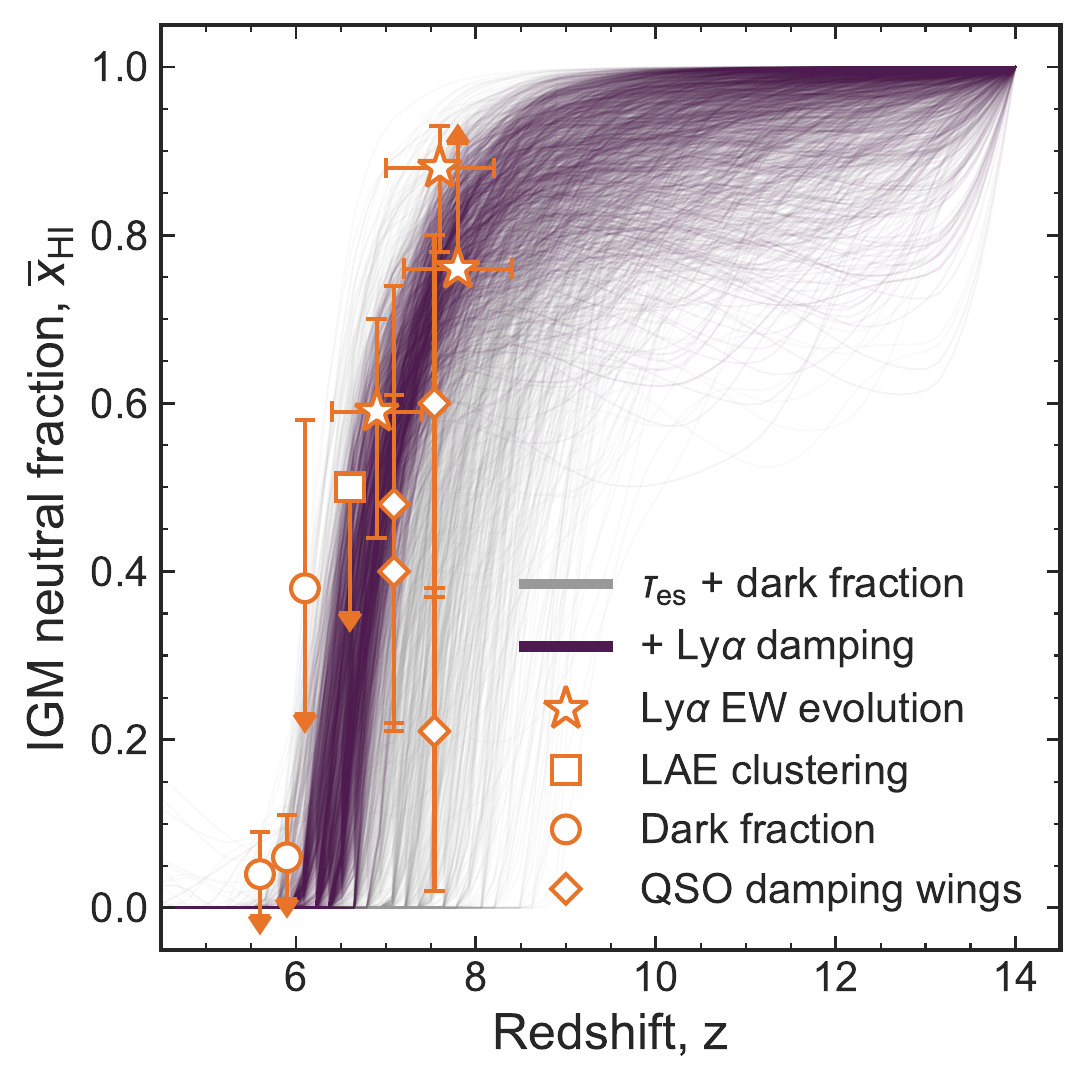}
    \vspace{-16pt}
    \caption{The volume-averaged IGM neutral hydrogen fraction as a function of redshift, $\xHI(z)$ from our inferred $\nion(z)$. Grey lines show 1000 draws from the posterior using only the CMB optical depth, $\tau_\mathrm{es}$ and dark fraction constraints, dark purple shows the inferred evolution using the additional observations, which prefers a later reionization. We also plot the constraints used in our likelihood (Section~\ref{sec:nion_like}).}
    \label{fig:history}
\end{figure}

\begin{figure}
	\includegraphics[width=\columnwidth]{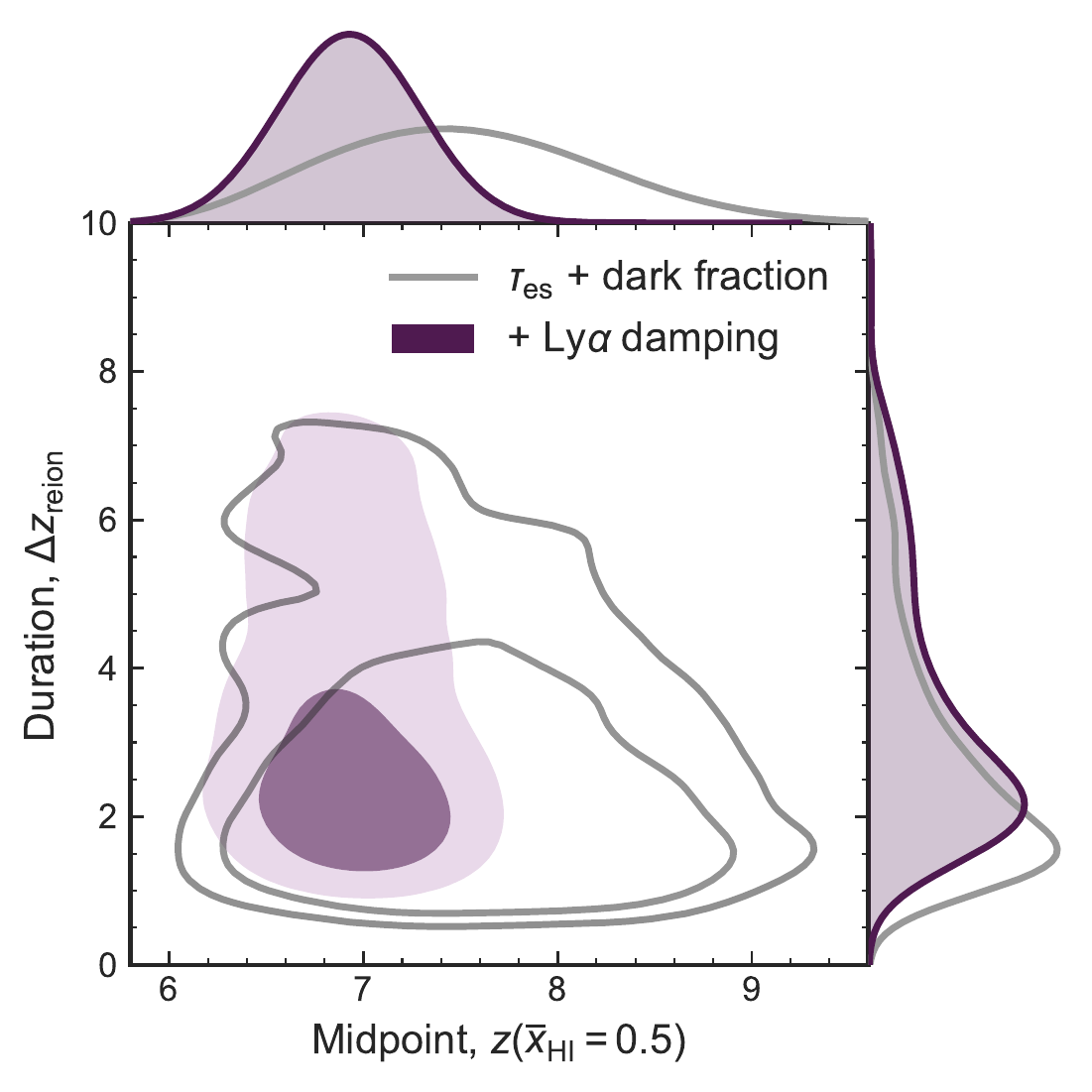}    
	\vspace{-16pt}
    \caption{1D and 2D distributions of reionization mid-point $z(\xHI = 0.5)$ and duration $\Delta z_\mathrm{reion}$ from our inferred $\nion(z)$. The grey contours shows the result using only the CMB optical depth, $\tau_\mathrm{es}$ and dark fraction constraints, dark purple shaded contours show the constraints using the additional \lya\ damping observations, which prefer a later reionization. Contours in the 2D plot show 68\% and 95\% confidence regions.}
    \label{fig:zreion}
\end{figure}

\begin{figure}
	\includegraphics[width=\columnwidth]{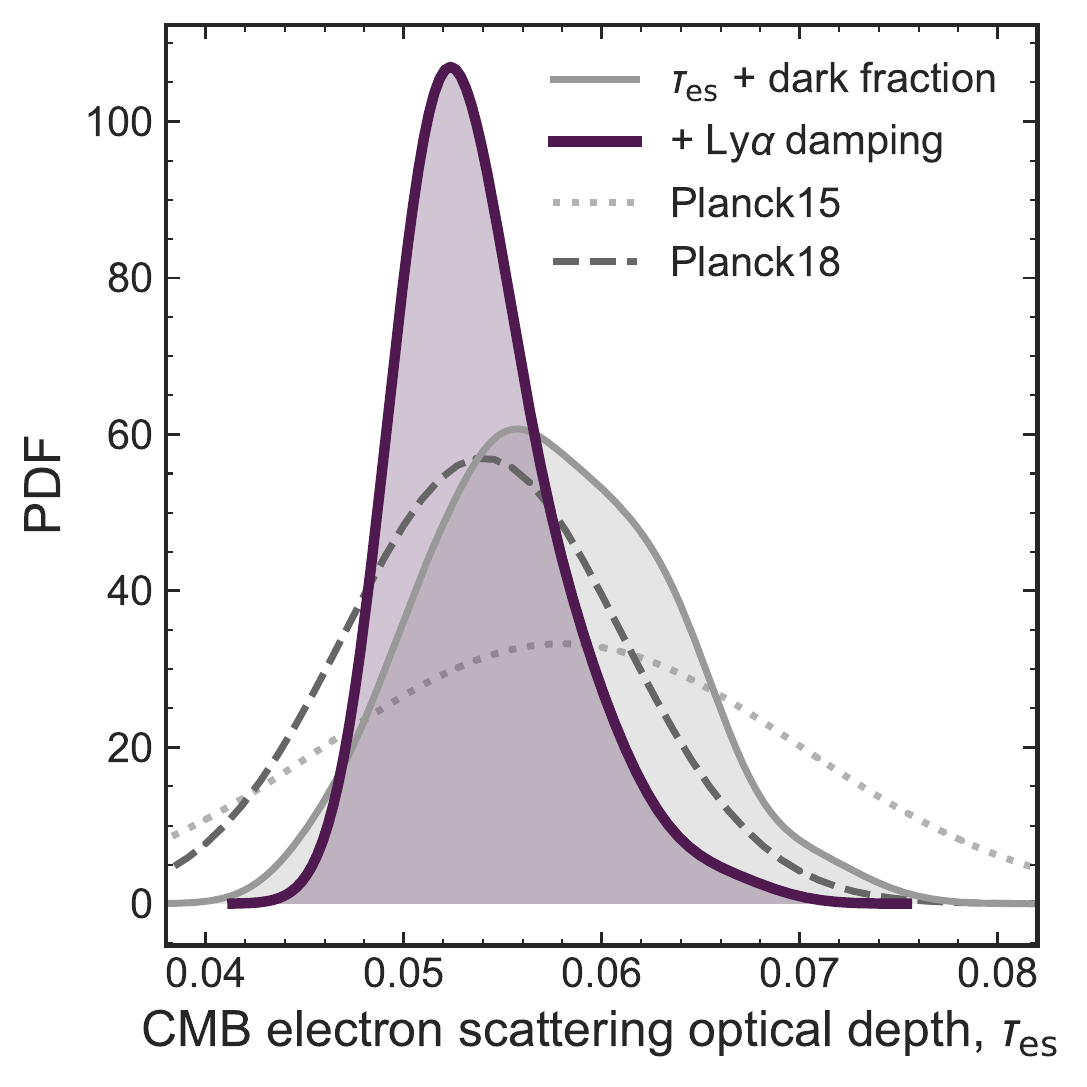}
    \vspace{-16pt}
    \caption{Electron scattering optical depth to CMB photons, $\tau_\mathrm{es}$. The \citet{PlanckCollaboration2015} and \citetalias{PlanckCollaboration2018} measurements are plotted in grey (light dotted, dark dashed respectively) assuming Gaussian errors. We plot the distribution of $\tau_\mathrm{es}(z=15)$ from our inferred $\nionbf$. As required to maximise our likelihood, $\tau_\mathrm{es}$ inferred from our model $\nionbf$ is fully consistent with the \citetalias{PlanckCollaboration2018} measurement (light grey distribution -- using the \citetalias{PlanckCollaboration2018} $\tau_\mathrm{es}$ and dark fraction, purple distribution -- including \lya\ damping constraints).}
    \label{fig:tau}
\end{figure}

Figure~\ref{fig:history} shows the reionization history obtained in our inference and the observational constraints used in the likelihoods (Section~\ref{sec:nion_like}). As required by the likelihoods our model fits the observations by construction, but it is instructive to observe the allowed reionization histories. While only using the most model-independent constraints allows a broad range of reionization histories \citep[see also][]{Greig2017c}, the \lya\ damping constraints prefer a later reionization.

The reionization timeline we infer is rapid between $z\sim6-8$, driven by the \lya\ damping measurements, and thus does not overlap significantly with an extended timeline, such as the model by \citet{Finkelstein2019}. This is primarily due to our sharply decreasing $\nion(z)$. In the \citet{Finkelstein2019} model low mass galaxies with extremely high escape fractions produce flatter $\nion(z)$ at early times than we infer (see panel (d) of Figure~\ref{fig:nion}), enabling reionization to start early, while more massive galaxies with lower escape fractions dominate the SFR density at later times, as the galaxy LF faint-end slope flattens, extending reionization. Our results imply that the latest reionization constraints favour an alternative balancing of the ionizing photon budget, likely reducing the dependence on very low mass galaxies with high $\fesc$ to complete reionization.

Double reionization (where the neutral fraction dips again at high redshift) is not ruled out \textit{a priori} in our model and can be seen in a few posterior draws. In most physical models the luminosity density of galaxies steadily declines at $z>10$ \citep[e.g.,][]{Mason2015,Tacchella2018a}, making double reionization unlikely. Within our framework, without placing model-dependent priors on the evolution of $\nion$, ruling out this scenario requires higher redshift measurements of the reionization timeline.

Figure~\ref{fig:zreion} shows the distributions of reionization's midpoint and duration obtained from our analysis. We define the midpoint: $z_\textrm{0.5} = z(\xHI=0.5)$, and the duration: $\Delta z_\mathrm{reion} = z(\xHI = 0.9) - z(\xHI = 0.1)$. With just the \citetalias{PlanckCollaboration2018} and dark fraction constraints, we obtain $z_\textrm{0.5} = 7.49_{-0.57}^{+0.71} \, (68\%)$ and $\Delta z_\mathrm{reion} = 2.26_{-0.99}^{+2.00}$, comparable to the \citetalias{PlanckCollaboration2018} analysis. Including the \lya\ damping constraints more tightly constrains the reionization midpoint to later times but does not significantly update its duration: $z_\textrm{0.5} = 6.93\pm0.14$ and $\Delta z_\mathrm{reion} = 2.83_{-0.99}^{+2.40}$. Again, this is due to the current lack of constraints on reionization's early stages ($z>8$).

Figure~\ref{fig:tau} shows the electron scattering optical depth to the CMB obtained in our inference. As required in our likelihood, both models are consistent with the \citetalias{PlanckCollaboration2018} $\tau_\mathrm{es}$ at $z=14$. Using the full $\xHI$ constraints we infer $\tau_\mathrm{es}=0.053\pm0.004$. 

\section{Discussion}
\label{sec:disc}

The CMB optical depth and dark fraction provide the most model-independent constraint on $\nion(z)$. Any model of reionization should produce $\nion(z)$ and reionization histories which are consistent with these results. As noted in Section~\ref{sec:results_nion}, the $\nion(z)$ we infer from these constraints is consistent with the decline in the galaxy UV luminosity density, with fixed $\fesc$ and $\xi_\textrm{ion}$, as shown in previous work \citep[e.g.,][]{Bouwens2015,Robertson2015}. As measurements of the reionization history improve, with further \lya\ observations and 21cm experiments, our knowledge of $\nion$ will increase and can be more directly decomposed into its constituent sources.

Figure~\ref{fig:nion} shows the \lya\ absorption constraints prefer a more rapid decline of $\nion$ over $z\sim6\rightarrow8$ to produce later reionization. While a contribution from quasars could be invoked to produce the build-up of $\nion$ from $z\sim8\rightarrow6$, using the model described in Section~\ref{sec:reion} the contribution of quasars to $\nion$ is negligible at $z>5$: the combined galaxies and quasar models shown in Figure~\ref{fig:nion} do not produce an evolution as steep as our inferred $\nion$. This is true even assuming quasars exist to $\MUV < -18$ and have hard ionizing spectra, $\alpha=-0.56$ \citep{Scott2004}. However, the faint-end slope of the quasar LF is still uncertain at $z>7$ \citep[e.g.,][]{Manti2016}: a large population of faint quasars at $z\sim8-10$ could contribute to the increase in $\nion$ at $z<8$.

The inferred rapid decline in $\nion$ is steeper than the evolution in the galaxy luminosity density (if integrated down to $\MUV \geq -13$). While many different effects contribute to $\nion$, using this luminosity density and constant $\fesc$ and $\xi_\mathrm{ion}$ it thus is difficult to describe $\nion$ evolution. Our inferred $\nion$ therefore provides tentative \textit{a posteriori} evidence that galaxies' ionizing photon emission properties evolve with redshift and/or that frequently-solicited, but undetected, faint galaxies contribute less to reionization than previously required \cite[e.g.,][]{Bouwens2015,Robertson2015}. 
For example, models where $\fesc$ is higher in more massive galaxies will naturally produce a steeper $\nion$ evolution, due to the more rapid build-up of high mass galaxies at high redshifts \citep{Sharma2018,Seiler2019,Naidu2019}. A steep $\nion$ can also be produced and/or enhanced if the galaxy luminosity density is dominated by `bright' galaxies ($\MUV \simlt -18$), which is the case if the UV LF faint-end slope is shallow, $\alpha > -2$. 
Accurate measurements of the faint-end slope at $z>4$ are therefore important for understanding the ionizing photon budget, and will be improved in the next decade with \JWST~deep imaging observations.
The dominant sources of reionization impact the topology of HII regions during reionization, measurable by future 21cm intensity mapping experiments: e.g. bright, massive sources produce more biased and larger ionized bubbles compared to if extremely faint, low mass sources dominate the ionizing budget \citep[e.g.,][]{McQuinn2007,Seiler2019}.

As demonstrated in Figure~\ref{fig:history} the \lya\ emitter clustering and EW evolution measurements most tightly constrain the $\nionbf$ posterior at $z\sim7$, due to their smaller uncertainties compared to the $\xHI$ measurements from individual quasars. As discussed in Section~\ref{sec:nion_like} the \lya\ damping measurements may introduce systematics due to modelling the intrinsic \lya\ emission (both in galaxies and quasars). In particular, an increase in circumgalactic medium absorption at $z>6$ could also play a role in absorbing \lya\ emission from galaxies \citep[e.g., from an increase in self-shielding systems][]{Bolton2013a}, but is likely subdominant to reionization as explaining the \lya\ damping without reionization requires the ionizing background to drop by at least a factor of 20 from $z\sim6\rightarrow7$ \citep{Mesinger2015}.

We find constraints from \lya\ damping provide increasing evidence that the mid-point of reionization was relatively late, $z\sim7$. However, the duration of reionization is still mostly unconstrained due to the lack of observations at $z>8$. Physical models of the $z>10$ galaxy population \citep[e.g.,][]{Mason2015,Tacchella2018a} predict a declining luminosity density, due to the lower abundance of halos to host star formation, which \textit{should} provide continually declining $\nion(z)$, but our work demonstrates that this is not yet confirmed observationally. $\nion$ is best constrained $z\sim6-8$, but future $z>8$ \lya\ surveys and 21cm experiments will measure $\xHI$, and thus $\nion$, at higher redshifts.

\section{Conclusions}
\label{sec:conc}

We have fit for the non-parametric evolution of the hydrogen-ionizing emissivity, $\nion$, at $z=4-14$, being maximally agnostic about the ionizing sources. We use the most recent constraints on hydrogen reionization to constrain our model. This method enables inference of $\nion(z)$ at redshifts higher than is possible with the \lya\ forest, which becomes too heavily absorbed at $z>6$. Our main conclusions can be summarised as follows:

\begin{enumerate}
    \item Current constraints on reionization favour a declining $\nion$ with redshift at $z>6$. This is moderately favoured by model-independent $\tau_\mathrm{es}$ and dark fraction constraints, $d\log_{10}\nionbf/dz|_{z=6\rightarrow8} = -0.31\pm0.35$ and more strongly favoured when \lya\ damping constraints are included ($d\log_{10}\nionbf/dz|_{z=6\rightarrow8} = -0.44\pm0.22$).
    \item $\nion(z)$ inferred from model-independent reionization constraints is consistent with the declining UV luminosity density of galaxies, with constant $\fesc$ and $\log_{10}\xi_\mathrm{ion}$, as found by previous studies.
    \item When reionization constraints from \lya\ damping in quasars and galaxies are included, a more rapid decline in $\nion(z)$ is inferred, which is less likely to be explained purely by the declining galaxy UV luminosity density (integrated to $\MUV > -13$) with constant ionizing photon output of galaxies, relative to their non-ionizing UV emission, $z=8\rightarrow6$.
    \item Including the \lya\ damping measurements significantly constrains the midpoint of reionization to $z_\textrm{0.5} = 6.93\pm0.14$ (compared with $7.49_{-0.57}^{+0.71}$ from the \citetalias{PlanckCollaboration2018} optical depth and dark fraction alone), but does not provide more information on its duration due to the lack of observational constraints in reionization's earliest stages, $z>8$.
\end{enumerate}

Without making assumptions about galaxy evolution, our analysis demonstrates current reionization measurements are broadly consistent with the evolution in the galaxy luminosity density, but hint at evolution in the ionizing photon output at $z>6$ and/or a lessened requirement for undetected extremely faint galaxies to dominate reionization. Future measurements of the reionization history, from galaxy surveys and 21cm experiments, can be used in this framework to more tightly constrain the evolution of $\nion$, and combined with statistics of the $z>8$ galaxy population, observable with \JWST, and measurements of the reionization topology, expose the sources of reionization.

\section*{Acknowledgements}

The authors thank Charlie Conroy for useful discussions and for providing comments on a draft of this paper. We thank Steve Finkelstein, Fred Davies and Brad Greig for useful discussions and for providing: the $\nion$ model from \citet{Finkelstein2019}, and $\xHI$ posteriors from \citet{Davies2018} and \citet{Greig2019} respectively. We thank Austin Hoag for sharing the $\xHI$ posterior from \citet{Hoag2019a}.
CAM acknowledges support by NASA Headquarters through the NASA Hubble Fellowship grant HST-HF2-51413.001-A awarded by the Space Telescope Science Institute, which is operated by the Association of Universities for Research in Astronomy, Inc., for NASA, under contract NAS5-26555. RPN gratefully acknowledges an Ashford Fellowship and Peirce Fellowship granted by Harvard University. ST is supported by the Smithsonian Astrophysical Observatory through the CfA Fellowship. JL is supported by an NSF Astronomy and Astrophysics Postdoctoral Fellowship under award AST-1701487.

\noindent
\textit{Software}: \verb|IPython| \citep{Perez2007a}, \verb|matplotlib| \citep{Hunter2007a}, \verb|NumPy| \citep{VanderWalt2011a}, \verb|SciPy| \citep{Oliphant2007a}, \verb|Astropy| \citep{Robitaille2013}, \verb|dynesty| \citep{Speagle2019}.


\bibliographystyle{mnras}
\bibliography{library} 



\appendix

\section{Impact of clumping factor}
\label{app:clumping}

\begin{figure}
	\includegraphics[width=\columnwidth]{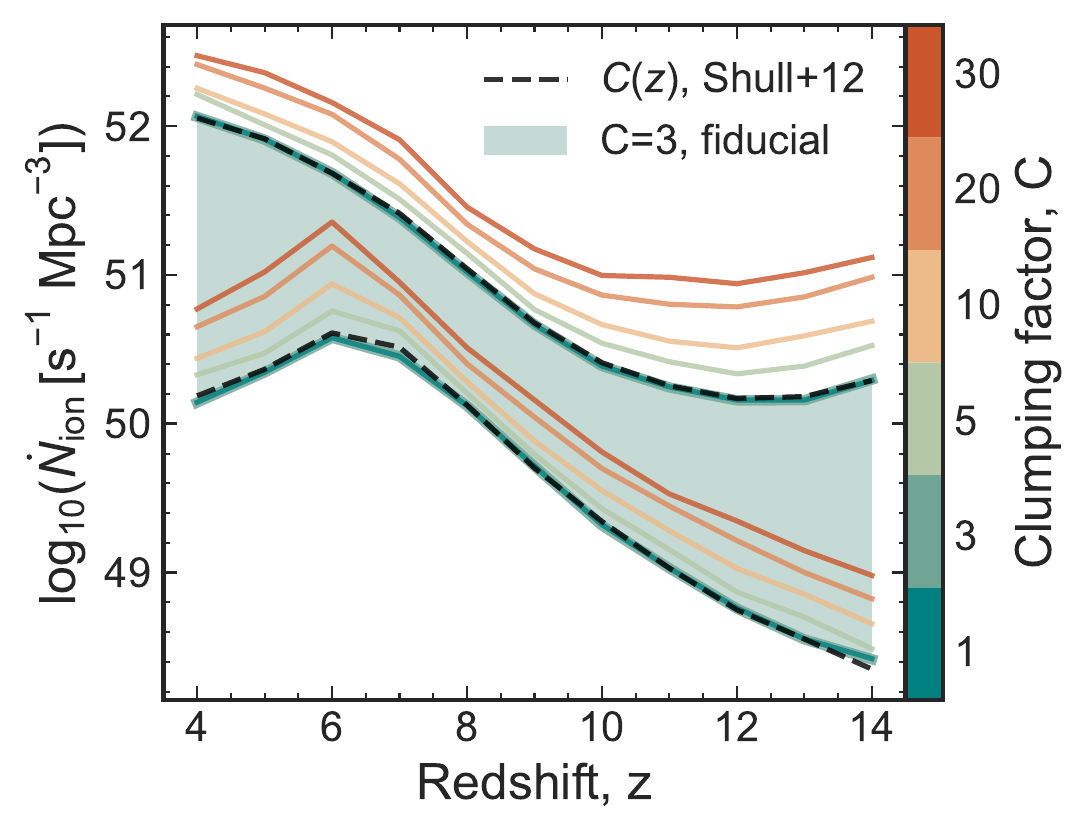}
    \vspace{-16pt}
    \caption{$\nion$ inferred from the \citetalias{PlanckCollaboration2018} $\tau_\mathrm{es}$ and dark fraction constraints, using different values of the clumping factor, $C$. Lines of the same colour bracket the 68\% confidence regions of the posterior $\nionbf$ at a given clumping factor (shown in the colourbar). Clumping factor values in the expected range ($C\sim1-5$) are shown in green, with the fiducial $C=3$ region shaded. Higher (and less likely) values shown in orange, which prefer higher $\nion$ to compensate for the faster recombination time in a clumpier IGM. We also show $\nionbf$ obtained using the evolving $C(z)$ model from \citet{Shull2012}. In all cases, the obtained $\nionbf$ are consistent within the 68\% region with our fiducial $C$ value.}
    \label{fig:clump}
\end{figure}

The only parameter in Equation~\ref{eqn:reion_Q} with significant uncertainty is the IGM clumping factor, $C$, which could impact our results. However, as noted by \citet{Bouwens2015}, $\nion$ is remarkably insensitive to $C$. We tested our inference with $C=1 - 30$ \citep[where $C=2-6$ is the physically motivated range based on IGM simulations, e.g.,][]{Finlator2012,Kaurov2015}, shown in Figure~\ref{fig:clump}. 

Increasing the clumping factor shifts the normalisation of $\nion$ to higher values, in order to balance increased recombinations in the IGM and complete reionization on time, but does not change its evolution with redshift. All of the resulting $\nionbf$ posteriors are consistent within the 68\% confidence region, and negligibly different for $C\sim1-10$. We also tested an evolving $C(z) = 2.9(1+z/6)^{-1.1}$ \citep{Shull2012} and found no significant difference from the non-evolving cases. Therefore, our inferred $\nionbf$ is robust to assumptions about the clumping factor, suggesting the impact of the clumping factor is less than the uncertainties in inferring $\nion$ from reionization constraints. All our results on the reionization timeline are insensitive to changes in the clumping factor as $\nion$ is modified self-consistently with $C$ to produce $\xHI(z)$ consistent with observations.

\bsp	
\label{lastpage}
\end{document}